# TARGET MULTIWIRE FOR THE FERMILAB BOOSTER NEUTRINO BEAMLINE

R. Prokop† Fermi National Accelerator Laboratory, Batavia IL, U.S.A.

*Abstract*

The Booster Neutrino Beamline experiment requested a new secondary electron emission multiwire profile monitor installation. The device had to be durable in high radiation conditions and mounted within a large 10-foot airtight steel fixture for installation near the beam target. Previous iterations of multiwire suffered radiation damage to both the connectors and wires. To ensure accurate horizontal and vertical beam profile measurements, an updated design was proposed, designed, and constructed. The new BNB multiwire utilizes 3 mil diameter gold-plated tungsten sense wires soldered to vertical and horizontal Alumina-96 ceramic planes, 50 wires per plane. Radiation hard Kapton insulated 30-gauge wires carry the output signals. Profiles are readout through charge integrator scanner electronics. This paper will detail the design and functionality of the BNB target multiwire and present relevant beam profile data.

## FUNCTION AND APPLICATION

The Booster Neutrino Beamline Multiwire for Horn 5 is a unique secondary emission monitor design at Fermilab utilized specifically for profile measurements of high intensity beam, up close to the experimental target [3]. This specialization calls for high durability and several modifications from the typical multiwire. The goal is to get clean gaussian profile measurements from small diameter wire, using wire/cable that can withstand high radiation environments. Previous iterations of this design used the classic multiwire flat Kapton ribbon cables with zero insertion force connectors glued to the ceramic frames. This is not ideal, as the intense environment of this device causes damage to the connectors. To remedy this problem, small gauge (30-40 AWG) Kapton insulated wire are individually cut to length and wrapped in bundles, soldered to both the frame and readout connectors. This presents many engineering challenges, but also opportunities to improve and craft a very durable and effective multiwire. This project has been a joint effort between the Fermilab Accelerator Directorate Instrumentation, External Beams & Target Systems groups.

The BNB multiwire operates on the principle of secondary electron emission, much the same as all other SEM profile monitors at Fermilab. The particle beam made up of protons strikes the sense wires and knocks off surface electrons from the metal material. This phenomenon leaves a resultant net positive charge on the wires which can be readout as a current through the signal cables. This current is integrated through transimpedance amplifiers in the front-end electronics and digitized as a voltage value to be plotted as a final beam profile measurement. The resultant signal is expected to be proportional to the amount of beam particles striking the wires [1].

Bias planes are comprised of titanium foils, and act to enhance the secondary emission effect by collecting stray electrons. The bias operates at +200VDC.

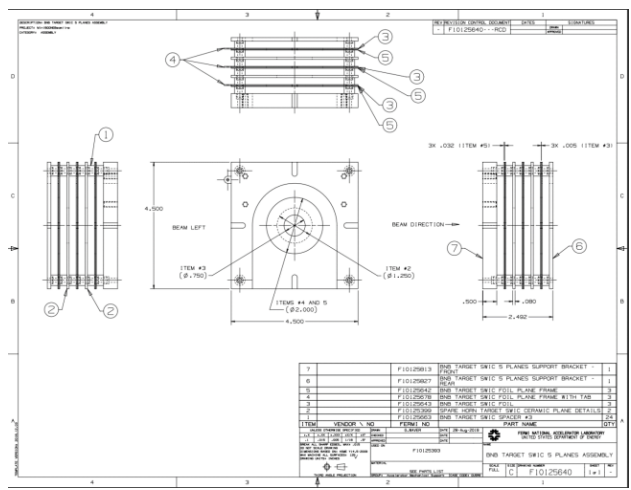

Figure 1: Multiwire assembly drawing.

## DESIGN

The BNB Target multiwire is an open air mounted secondary emission monitor designed for the high intensity beam readout. The multiwire assembly as shown in Figure 1, consists of two ceramic planes, one of vertical orientation and one horizontal. The ceramics are 4.5 x 4.5 inch with a center hole of diameter 1.25 inch. They have 50 pads at 1mm spacing for signal wires, and 100 pads at 0.5mm spacing for sense wires across the center hole. There are three titanium foils of .005-inch thickness, which act as bias planes to assist in maintaining an electric field for charge deposition on the sense wires. The multiwire assembly is held together by front and back support brackets as well as spacers between the planes. The multiwire mounting assembly is slid into a larger horn mechanical enclosure for insertion close to the target. Discussion of the entire mechanical mounting assembly is beyond the scope of this document.

The signal carrying conductors are fashioned in a similar way to previous the MiniBooNE target multiwire, but without the zero insertion force connectors or Kapton ribbon cables. This new design instead utilizes 100 individual Kapton insulated wires of 0.25mm diameter for the signal wires, terminated at one end at the pads of the ceramic and at two 50 pin D-Sub connectors on the other end. The use of connectors and epoxy was not ideal for this application, as in previous designs the epoxy wore down and connectors became damaged due to heat and radiation. 0.003-inch diameter AuW wires are soldered pad to pad for the center beam sense wires. AuW wires provide acceptable signal integrity through collected charge, as well as having excellent density and durability with high intensity beam [2]. A small diameter of 3 mil is typical for applications such as this; it limits the amount of material in the beam while also giving ample signal.

---

\* This work was supported by the U.S. Department of Energy under contract DE-AC02-07CH11359

† rprokop@fnal.gov

*Wire Preparation*

- The Kapton insulated wires make for a challenge when assembling the multiwire. To account for every sense wire reading, there must be 50 wires per plane, totalling to 100 signal wires. Additionally, there is one high voltage wire which supplies the bias. The signal wires for the multiwire must measure at minimum 10-feet in length. When accounting for some slack, they were cut to ~11 ft. The wires must reach from the multiwire planes to the D-Sub connectors for readout. To achieve a neat and organized bundle of wires, the decision was made to separate the wires into sets of 10. Each plane has five sets of 10 wires, with each set taped together at intervals along the 10-foot length run. The mechanical assembly, shown in Figure 2, has a 9-foot metal pipe which carries the wire bundles from the wireplane mount to the connectors. There is a slight angular bend at the end of the pipe, which makes the pipe length and wireplane fixture distance approximately 9.5 ft.
- The ends of the Kapton wires are stripped to reveal a few mm length of centre conductor for soldering. It is important to consider the order of the wires when preparing bundles. Each set of 10 wires has a red tick mark on the electrical tape near the first wire. A multimeter was used to check continuity between ends of the wires and ensure that each wire is properly marked on both sides.

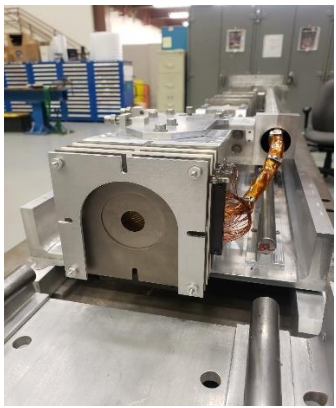

Figure 2. Multiwire assembly completed and mounted.

*Soldering Procedure*

The soldering of signal wires requires a pre-made wire containing fixture, as a procedure like this has never been attempted before with a multiwire. The signal wires are soldered to pads directly on the ceramic itself, much different than other modern multiwires which have connectors and 50 conductor cables or ribbon cables. The fixture is simply a metal block, with 0.25mm width slots to sit the wires in. A top block is fastened above the wires to clamp down and secure them in place. This wire fixture is essential for the completion of the multiwire assembly.

The central sense wires are soldered with an intricate method by a wireplane specialist at Fermilab. She winds the wires on a transfer plane to get tension and the correct spacing, 0.5mm in this case. They are aligned to the pads on the ceramic thereafter. Then the wires are temporarily epoxied to tape at the end of the substrate and soldered.

To circumvent damaging the ceramic with high heat, and allow for fast and easier reflow soldering, Alpha CVP-520 low temp solder paste [4] was initially chosen as the bonding agent. This paste has a melting point of 138ºC and an optimum reflow temperature between 150ºC and 190ºC. A soldering iron would work for drag soldering but runs the risk of applying too much focused heat to the ceramic. A heat gun causes far too much spread, with risk of affecting the sense wires or too much airflow. The ideal tool for the job is an adjustable airflow heat wand with changeable tips. Keeping airflow relatively low allows more heat with less risk of paste blowing across the ceramic. A temperature of 190ºC/374ºF provides a fast and effective reflow for the solder paste and bonds the wires efficiently.

The 10-foot bundles were routed through 9-foot pipe to the readout box at MI-8 during the full assembly procedure. The readout cable and connector soldering process involved several days and many hours of stripping and soldering each wire end to the D-Sub connectors bolted to the readout box. A high voltage wire was also connected to the planes and fed to an SHV bulkhead connector.

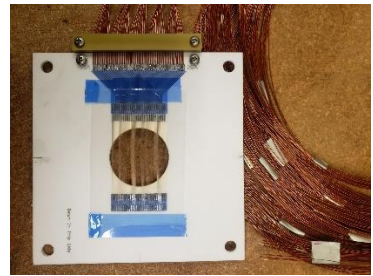

Figure 3: Wireplane with Kapton insulated signal wires.

## INSTALLATION AND RESULTS

The multiwire is mounted on a mechanical assembly and aligned with precision instrumentation. This process ensures that the center of the wireplane grids and the center of the beam pipe are within ~1mm tolerance. Kapton insulated cabling is routed through a 9 ft stainless steel pipe for protection. At the end of the pipe is a welded cap with D-Sub connector feedthroughs and one SHV-5 high voltage connector feedthrough. The signal wires have been carefully soldered individually to each pin and tested for continuity as well as shorts.

The mounting assembly is on rails, and during installation it slides into a larger housing unit. This unit is sealed completely airtight and placed near the beam target. Due to the proximity to the beam target and the high radiation hazard, access to the internals of the unit for repairs or modifications is very difficult and time consuming. All alignment and testing are done to the highest of standards to reduce the risk of failures.

*Readout Electronics*

The BNB target multiwire uses standard Fermilab readout electronics for beam profile monitors. These chassis are called scanners, shown in Figure 4. A set of boards housing 96 channel analog front end integrator circuits receive the signals from the current carrying cables and integrate them to voltages for digitization. Wire 1 and 50 on the vertical and horizontal wireplanes act as ground wires, tied to chassis ground at the scanner. The modern scanners have ethernet capability, for fast and reliable control and data capture. Every scanner is connected to the Fermilab Accelerator Control Network front end for viewing data and making changes to settings. The Fermilab Controls Department manages these devices and provides Instrumentation with reliable readout electronics.

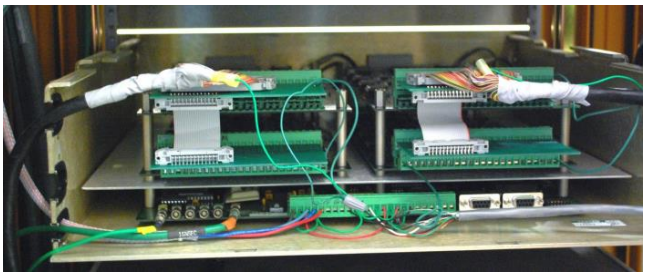

Figure 4: Rear view of scanner electronics.

*Profile Data*

The data pictured below is not from the BNB target multiwire, as this device is recently installed and has not yet seen beam. The image below shown in Figure 5 is a screen capture from a similar device, a standard multiwire SEM. This is the horizontal and vertical display showing expected results with a Gaussian fit. The histogram plot has each wire voltage value represented as a blue bar. The center of beam is ideally the center peak of the profile.

The environment which the BNB target multiwire is installed is considered a high radiation area, due to its proximity to the beam target. Exact radiation measurements are not known but are enhanced by secondary beam materials such as neutrons. The beam bunch intensity is approximately 4E12 protons/bunch and the proton rate is around 3E16 protons/hr at a proton beam energy of 8 GeV.

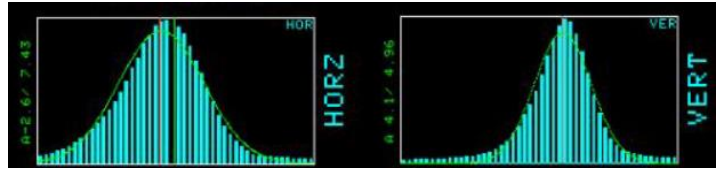

Figure 5: Beam profiles, horizontal and vertical

*Future*

The readout electronics consist of two flat-to-round signal cables, vertical and horizontal, connected to a Controls Scanner rack-mount chassis. The readout electronics are housed at Main Injector Service Building 12. Testing the device is possible using the common flash test procedure for multiwire profile monitors. This test uses an integrating scanner, high voltage supply, and video monitor to view the charge collection as a histogram of voltages on each wire. Due to the nature of the BNB multiwire and it's very difficult to reach location, as well as nearby hazards, it is challenging to repair. A complete failure or wire replacement would result in the entire assembly's removal and a disassembly of the wireplane fixture.

Another Target multiwire for BNB Horn 4 is coming, and there are lessons learned and proposed modifications the team has come up with. For the signal wire readout cables, there is a ceramic coated small gauge wire produced by a vendor which may be a suitable substitute for the Kapton insulated wire. The ceramic coated wire is highly durable and highly flexible, while maintaining its form. The ceramic coating is extremely small, thin, and porous. The primary challenge when implementing this material will be stripping the coating from the central Copper/Nickel conductor. Using a molten hot solution is one method, but highly hazardous and not ideal for 100 wires. Fermilab engineers who have done research this potential wire type, have proposed a few alternate stripping and mounting methods which will be tested.

## CONCLUSION

The BNB target multiwire was a collaborative effort between multiple departments and teams at Fermilab. It overcame numerous challenges during design and assembly. The target systems group expects full operation during the 2023-24 accelerator run, with hopes for quality profile data to follow. The radiation-hard connectors and wire strain relief, as well as the Kapton insulation should provide this multiwire with many years of safe operation within the BNB target enclosure. Lessons learned from this project will be carried over into future installations.

## ACKNOWLEDGEMENTS

Thanks to the following past and present Fermilab team members who lent their knowledge, skills, and time to this project. Daniel Schoo, retired Fermilab Electrical Engineer. Gianni Tassotto, retired Fermilab Engineering

Physicist. Thomas Kobilarcik, Fermilab Engineering Physicist. George Lolov, Fermilab Mechanical Engineer. Ming-Jen Yang, late retired Fermilab Physicist. Stephen Rocos, Fermilab Technician. Wanda Newby, Fermilab Technical Specialist. Paula Lippert, Fermilab Technical Specialist.## REFERENCES

[1] G. Tassotto, J. Zagel, "Beam Profile Detectors at the New Fermilab Injector and Associated Beamlines", in Proc. DIPAC, Chester, UK, 1999, pp. 141-143.
[2] Hsing-yin Chang, Andrew Alvarado, Jaime Marian, "Calculation of secondary electron emission yields from low-energy electron deposition in tungsten surfaces", Applied Surface Science, Vol. 450,
[3] Fermilab, https://ad.fnal.gov/ebd/
[4] Alpha, https://www.macdermidalpha.com/assembly-solutions/products/solder-paste/alpha-cvp-520-solder-paste